\documentclass{aastex}
\usepackage{emulateapj5}
\usepackage{graphicx}
\usepackage{times}
\newcommand{\Ni}{\ensuremath{^{56}\mathrm{Ni}~}}
\newcommand{\Coa}{\ensuremath{^{56}\mathrm{Co}~}}
\newcommand{\Cob}{\ensuremath{^{57}\mathrm{Co}~}}
\newcommand{\Ti}{\ensuremath{^{44}\mathrm{Ti}~}}

\newcommand{\ergss}{ergs$\;\mbox{s}^{-1}$}

\makeatletter

\newenvironment{inlinefigure}{%
\def\@captype{figure}%
\noindent\begin{minipage}{0.999\linewidth}\begin{center}}
{\end{center}\end{minipage}\smallskip}

\newenvironment{inlinetable}{%
\def\@captype{table}%
\noindent\begin{minipage}{0.999\linewidth}\begin{center}\footnotesize}
{\end{center}\end{minipage}\smallskip}

\begin{document}

\submitted{ApJ in press}
\title{Rates of Observable Black Hole Emergence in Supernovae}

\author{Shmuel Balberg\altaffilmark{1} and Stuart L.~Shapiro\altaffilmark{2}}

\affil{Department of Physics, Loomis Laboratory of Physics,
University of Illinois at Urbana--Champaign, 1110 West Green Street,
Urbana, IL 61801--3080}

\altaffiltext{1}{Permanent address: Racah Institute of Physics,
The Hebrew University, Jerusalem, Israel 91904, 
{\tt shmblbrg@saba.fiz.huji.ac.il}}

\altaffiltext{2}{also Department of Astronomy and National Center for
Supercomputing Applications, University of Illinois
at Urbana--Champaign, Urbana, IL 61801, 
{\tt shapiro@astro.physics.uiuc.edu}}

\begin{abstract}

A newly formed black hole may be directly identified if late-time accretion of 
material from the base of the ejected envelope generates a luminosity 
that is observable in the tail of the supernova light curve. In this work we 
estimate the rate at 
which events where the black hole ``emerges'' in the supernova light curve can 
be detected with present capabilities. Our investigation is based 
on an analytical model of the accretion luminosity at emergence as a function 
of progenitor mass, coupled to the inferred rate of observed Type II 
supernovae in nearby galaxies. We find through a parameter survey that under 
optimistic assumptions the potential rate of observable events can be as 
high as several per year. However, supernovae which produce black holes are 
also likely to be low energy explosions and therefore subluminous, as was 
the case for the best candidate to date, SN1997D. If black hole-forming 
supernovae are underdetected owing to lower luminosities, the rate of 
observing black hole emergence is probably not larger than once every few 
years. We therefore emphasize the importance of dedicated searches for nearby 
supernovae as well as faint supernovae projects for improving the 
prospects of observationally certifying the supernova--black hole connection. 

\end{abstract}


\section{Introduction}\label{Sect:Intro}

Theoretical studies of core-collapse supernovae suggest that the remnant they 
leave behind can be either a neutron star or a black hole. The nature of the 
remnant is determined mostly by the amount of material the collapsed core 
accretes while the explosion is still progressing: if the final mass of the 
core exceeds the maximum mass it can sustain, it will collapse to a black hole 
rather than evolve to a stable neutron star.  

Observational support for this theoretical scenario is not complete.
Radio pulsar emission and x-ray observations provide observational 
evidence associating neutron stars with sites of known supernovae, but similar 
evidence for a black hole - supernova connection is still mostly 
unavailable. Indirect evidence that the black hole 
candidate in the x-ray binary system
GRO J1655-40 was formed in a supernova explosion has recently 
been reported by \citet{Israelianal99}. They base their inference on the 
relatively high abundances of nitrogen and oxygen on the surface of 
the companion, which has too low a mass to have produced such abundances by 
thermonuclear burning. Assuming those elements were deposited on the 
companion by the supernova explosion of the primary star provides indirect  
evidence of a black hole formed in a supernova. 

In a series of recent papers we examined the prospects of directly identifying 
a black hole created in a supernova through late-time accretion onto it 
\citep{ZCSW,SN1997DLet,BHinSN}. Such accretion is powered by material  
from the base of the ejected envelope which remains bound to 
the black hole and gradually falls back and is accreted.
Analytic \citep{CSW96} and numerical \citep{ZCSW} investigations showed 
that the rate of late-time spherical accretion, when it becomes dust like, 
is expected to decline as a power law in time and so is the corresponding 
accretion luminosity. Such an accretion luminosity will produce a distinct 
signature on the total light curve if and when it becomes comparable to 
the output of other power sources, i.e., the initial internal energy of the 
envelope and decays of radioactive isotopes synthesized in the explosion. 
If detected, the unique time dependence of the accretion luminosity 
due to fall back would be a direct sign of the black hole ``emerging'' in the 
supernova light curve. 

To date, black hole emergence in supernovae is yet to be observed. The 
prominant obstacle is that radioactive heating in the light curves of 
typical core collapse supernovae obscures any 
accretion luminosity for very long times. Even though radioactive heating 
decays exponentially while accretion luminosity declines only as a power law, 
the absolute power in radioactive heating remains the dominant source 
of the light curve until the supernova is no longer detectable. The case of 
SN1987A is a useful example \citep{ZCSW}: while no firm evidence of a newly 
formed neutron star has been found, positive indication of the presence of a 
black hole (if one was produced) would be unobservable due to radioactive 
heating for about 1000 years. 

Viable candidates for observing black hole emergence are 
supernovae which show very little radioactive heating. In principle, larger 
mass remnants, which are more likely to be black holes \citep{Fryer99}, are 
also more likely to be associated with supernovae which show diminished 
abundances of radioactive isotopes. Radioactive isotopes are 
produced in the deepest part of the exploding star, and therefore can be 
absorbed in a larger remnant in the earliest stages of fallback 
\citep{WoosWeav95}. The best two recent candidates for 
supernovae which ejected very little radioactive 
isotopes and possibly produced black holes were SN1994W and SN1997D. 
In SN1997D the abundance of \Coa inferred in the light curve was only 
$2\times 10^{-3}\;M_\odot$, and the explosion energy was estimated as only 
$4\times 10^{50}\;$ergs, both significantly lower than typical in Type II 
supernovae \citep{SN1997D}. Our previous analyses \citep{SN1997DLet,BHinSN} 
focused on SN1997D and showed that a $3\;M_\odot$ black hole in its remnant  
may emerge about 1000-1200 days after the explosion. Unfortunately, 
even at emergence the luminosity would have been 
only marginally detectable with HST. In the case of SN1994W 
$2\times 10^{-3}\;M_\odot$ of \Coa was the upper limit for radioactive 
heating, so the accretion luminosity might have emerged much earlier 
(and at higher value than in SN1997D). Unfortunately (again) the tail of the 
light curve was apparently dominated by circumstellar interaction 
(which was absent in SN1997D), and thus obscured any weaker source of heating 
from inside the envelope \citep{Sollerman94W}. 

Can we expect better success in a systematic quest for 
observing black hole emergence in supernovae? Our goal here is to 
perform a first estimate of the potential rate of observations, and examine 
possible strategies for improving the prospects of such an observational 
search. We base our estimate on an analytic approximation for the luminosity 
at emergence of a black hole in a supernova as a function of progenitor mass, 
and then couple it to estimates of the rate of supernovae in 
nearby galaxies. Although the model is crude and excludes several details 
which determine specifics of supernovae explosions, their light curves and 
remnant formation, we are able to point to some underlying features. 
Our conclusions suggest 
that the likely rate at which black hole emergence can be observed with 
present capabilities and supernovae searches is probably once every several 
years, although with favorable assumptions this rate can be as high as about 
once a year. We also find that the dominant factor in determining this rate 
is most likely the difficulty of detecting nearby faint supernovae. A 
systematic search for such supernovae would significantly enhance the 
probability of discovering black hole emergence and observationally 
confirming the black hole - supernova connection.

We begin in \S~\ref{Sect:Model} where we develop the model that relates the 
emergence luminosity to progenitor mass. This function serves as the 
basis for estimating the values of emergence luminosities, presented in 
\S~\ref{Sect:LvsM}. The main results are given in 
\S~\ref{Sect:RATES}, where the emergence luminosities are coupled to a 
distribution of progenitor masses, allowing for estimates of the rate of 
observable supernovae in nearby galaxies and providing us with an estimate 
of the rate of observable black hole emergence events. Conclusions and 
discussion are given in \S~\ref{Sect:End}.

\section{The Model}\label{Sect:Model}

To assess the observability of black hole emergence we must first determine 
the accretion luminosity of the black hole and its value at emergence. 
We employ a simplified model in which the emergence 
luminosity, $L_{BH}$, is a unique function of progenitor mass, $M_*$. 
This is clearly a crude approximation, since explosion energy, 
black hole mass and post-shock evolution are likely to be dependent on other 
factors, such as progenitor rotation \citep{FryHeg99}, metallicity 
\citep{WoosWeav95} and effects of a binary companion \citep{WillLang99}. 
Nonetheless, such an approach is a useful first approximation 
for deriving the observable event rate (and indeed is frequently used in other 
applications regarding core collapse supernovae, such as determining the 
neutron star and black hole initial mass functions; see 
Timmes, Woosley \& Weaver 1996, Fryer 1999, Fryer \& Kalogera 1999). 
We discuss some of the limitations and uncertainties concerning the unique 
function $L_{BH}(M_*)$ towards the end of this section.

\subsection{The Accretion Luminosity}

Constructing the function $L_{BH}(M_*)$ requires the time-dependence of 
the luminosity due to accretion onto the black hole. As shown analytically by 
\citet{CSW96} and numerically by \citet{ZCSW}, at late times after the 
explosion, {\it spherical} accretion onto the black hole becomes dust-like, 
and the accretion rate, $\dot{M}$, declines as 
$\dot{M}\propto t^{-5/3}$. \citet{ZCSW} also showed that the 
resulting accretion luminosity is consistent with the \citet{Blondin86} 
estimate of hypercritical accretion, where $L_{acc}\propto \dot{M}^{5/6}$, so 
that 
\begin{equation}\label{eq:L_accoft} 
L_{acc}\propto t^{-25/18}\;.
\end{equation}
This is a fundamental prediction of spherical accretion onto a black hole 
following a supernova, and can serve as distinct signature of accretion in 
the overall light curve \citep{ZCSW,BHinSN}. We address the question of 
likelihood of spherical accretion later in this section.    

The proportionality constant in equation (\ref{eq:L_accoft}) is dependent 
on the black hole mass, $M_{BH}$, and on the properties of the bound material, 
including composition, density and velocity profiles. For the purpose of an 
analytic estimate an effective time, $t_{dust}$, can be defined so that 
\begin{equation}\label{eq:L_acctdust}
L_{acc}(t)=L_{Edd}\left(\frac{t}{t_{dust}}\right)^{-25/18}\;
\end{equation}
where $L_{Edd}=4\pi c G M_{BH}/\kappa$ is the Eddington luminosity for 
material with opacity $\kappa$ accreting on a black hole of mass 
$M_{BH}$. For bound material which initially has a uniform density, $\rho_0$, 
and a homologous velocity profile, $v(r,t=0)=r/t_0$ (where $t_0$ is the 
{\it expansion time scale}), the time $t_{dust}$ can be roughly 
estimated as (Balberg et al.~2000, eq.~[30])
\begin{eqnarray}\label{eq:whatistdust}
  t_{dust} & = & 5.675 \left(\frac{\mu}{0.5}\right)^{-4/5}
                 \left(\frac{\kappa}{0.4}\right)^{3/10}
                 \left(\frac{M_{BH}}{M_\sun}\right)^{-1/5}\\ \nonumber
 & &  \left(\frac{\rho_0 t_0^3}
          {10^9\;\mbox{gm}\;\mbox{cm}^{-3}\;\mbox{s}^3}\right)^{1/2}\;
                                                             \mbox{days}\;,
\end{eqnarray}
where $\mu$ is the mean molecular weight of the bound material. This estimate 
for $t_{dust}$ includes the effect of radiation pressure during the earliest 
phase of fallback, when an accretion luminosity at about the Eddington limit 
modifies the flow of the bound material.
This seems to be the case for realistic supernova envelopes 
(see Balberg et al.~2000 for details). We emphasize that the time $t_{dust}$ 
is an effective quantity: at that actual time after the 
explosion the accretion is not expected to have settled 
into dust-like motion. Rather, $t_{dust}$ corresponds to the time when 
the luminosity {\it extrapolated back from late-time} 
equals the Eddington limit. 

\subsubsection{Explosion Energy and Black Hole Mass}

Equation (\ref{eq:whatistdust}) introduces the quantities required 
for estimating the accretion luminosity. For a given progenitor,
these quantities are determined primarily by the 
explosion energy and the manner in which it is distributed in the envelope. 
We follow the approach of \citet{FryKal99}, and parameterize the 
explosion by the total energy available for the 
through core collapse, $E_{exp}$, and the fraction, $f$, of this 
energy which is spent on unbinding the star. The remainder of the explosion 
energy is converted mostly to kinetic energy of the ejected material. 

The energy available to unbind the star can be combined with the binding 
energy profile of the progenitor to calculate the remnant mass, $M_{rem}$, 
as a function of $M_*$ according to
\begin{equation}\label{eq:M_rem}
f\times E_{exp}=\int_{M_{rem}}^{M_*} E_{BE}(m) d\!m\;,
\end{equation}
\begin{inlinefigure}
\centerline{\includegraphics[width=1.0\linewidth]{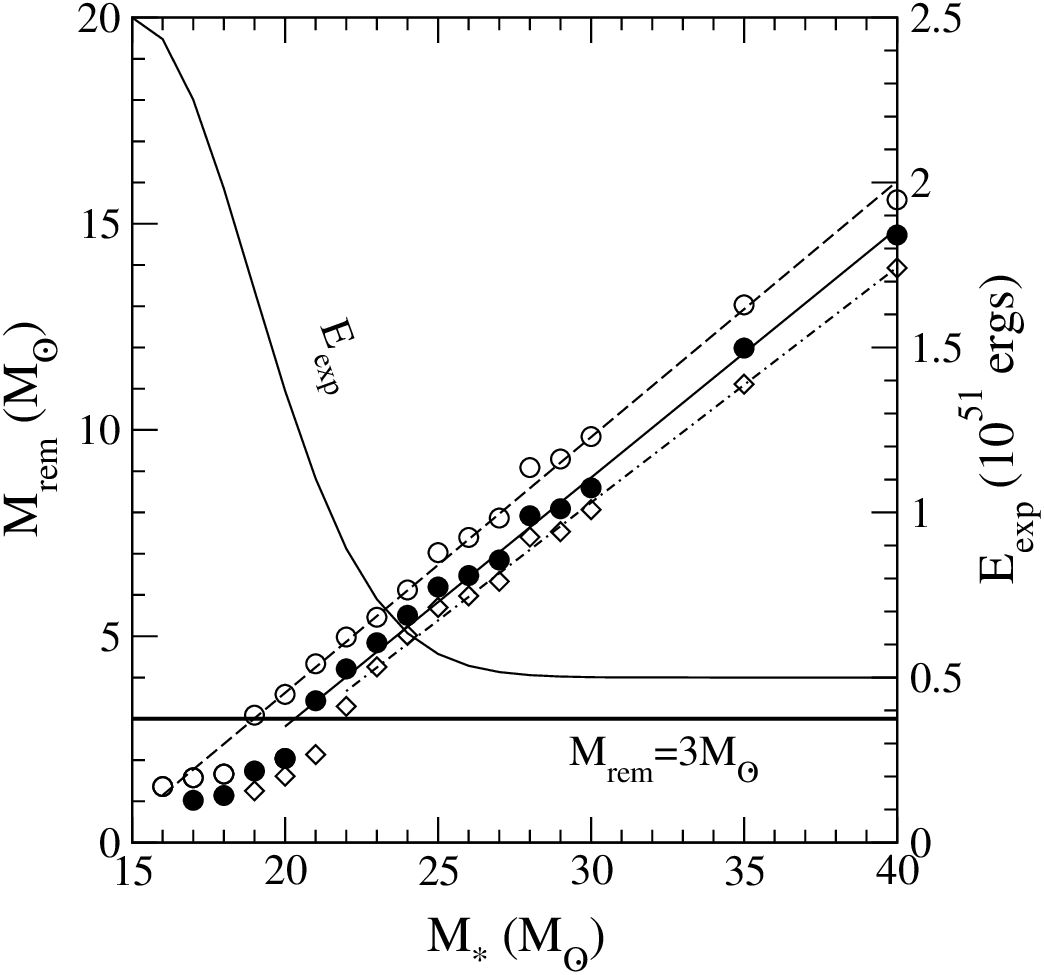}}
\figcaption{
Estimate of the total explosion energy, $E_{exp}$ (eq.~[\ref{eq:E_exp}]) and 
predicted remnant mass, $M_{rem}$ as a function of progenitor mass, $M_*$, 
for different values of the parameter $f$ (top to bottom: 
$f=0.25, 0.50$ and 0.75). Data points correspond to the values from the 
numerical model (eq.~[\ref{eq:M_rem}), lines are linear best fits for the 
$M_{rem}(M_*)$ function 
for high mass remnants. The solid line at $M_{rem}=3\;M_\sun$ marks the 
assumed threshold mass for black hole formation.\label{fig:M_BH_M_*}}
\end{inlinefigure}

\noindent
where $E_{BE}(m)$ is the binding energy profile, for which we use the massive 
star models of \citet{WoosWeav95}.
Figure~\ref{fig:M_BH_M_*} shows the remnant mass versus progenitor mass 
relation found with $f=0.25$, 0.5 and 0.75 where we adopted an analytic 
explosion energy function of
\begin{equation}\label{eq:E_exp}
E_{exp}=10^{51}\times
 \left[0.5+2\exp\left(-\frac{(15-M_*)^2}{30}\right)\right]\;\;\mbox{ergs}\;,
\end{equation}
for $M_*$ in units of $M_\sun$. This energy function (also plotted in 
Fig.~\ref{fig:M_BH_M_*}) is based on a fit to 2D core-collapse simulations 
of progenitors in the range $8\!\leq\!M_*\!\leq 40\;M_\sun$, evaluated as 
the change in energy before collapse and after about 1 second 
of the material that experiences explosive shock heating. 
\citep{Fryer99}. Note that for the higher range of progenitor masses, which 
are the candidates for forming black holes, the remnant mass 
varies about linearly with the progenitor mass. Obviously, the remnant mass 
increases with decreasing $f$. 

The nature of the remnant depends on whether its mass exceeds the 
maximum mass of a neutron star. As a conservative lower limit 
we will require $M_{rem}\geq 3\;M_\sun$ (baryonic mass, which is that found in 
eq.~[\ref{eq:M_rem}]) for the remnant to be a black hole. This consistent is 
with most realistic equations of state (see Cook, Shapiro \& Teukolsky 1994). 
  
\subsubsection{Properties of the Bound Material}

The estimate of the accretion luminosity includes a dependence on both 
the composition and kinetic energy of the bound material. For a homologously 
expanding medium the combination 
$\rho_0 t_0^3$ can be expressed in terms of the global quantities of the 
post-shock helium-rich layer, which is the source of the late-time accretion 
\citep{BHinSN}:
\begin{eqnarray}\label{eq:rho0t03}
\rho_0({\rm He}) t_0^3 & = 0.222\times 10^9 \left(\frac{M_{\rm He}}
{M_\odot}\right)^{5/2} & \\
& \left(\frac{E_{kin}({\rm He})}{10^{49}\mbox{erg}}\right)^{-3/2}\;
\mbox{gm}\;\mbox{cm}^{-3}\;\mbox{s}^{-3} & \;\;, \nonumber
\end{eqnarray}
where $M_{He}$ is the mass of the layer and $E_{kin}({\rm He})$ is its total 
kinetic energy. 

The mass of the helium layer can be estimated by subtracting the remnant 
mass calculated in equation~(\ref{eq:M_rem}) from the mass of the helium core 
of the progenitor. We parameterize the kinetic energy of the helium layer as a 
fraction, $f_{\rm He}$, of the total kinetic energy delivered to the ejected 
envelope: $E_{kin}({\rm He})=f_{\rm He}(1-f)E_{exp}$. Simulations of 
Type II explosions typically suggest that $f_{\rm He}$ is of the order of a 
few percent \citep{Nomotoal94,ArnettBook}, generally decreasing for larger 
mass progenitors. The exact value does, however, depend on the details of the 
explosion and on progenitor structure. 
 
\subsection{Emergence Time and Luminosity}

The black hole mass and the effective time $t_{dust}$ uniquely determine  
the time dependent accretion luminosity (eq.~[\ref{eq:L_acctdust}]) which can 
then be compared to other sources of emission in the supernova.
Specifically, we define the time of emergence, $t_{BH}$, as that when the 
accretion luminosity reaches $\frac{1}{2}$ of the total bolometric  
luminosity. This time and the corresponding emergence luminosity, 
$L_{BH}\equiv L_{acc}(t_{BH})$, depend on 
the source of luminosity with which accretion competes. This source is either 
the initial internal energy deposited in the envelope by the explosion, or 
ongoing heating due to decay of radioactive isotopes.  

The radioactive isotopes relevant to powering the light curve, 
\Ni and its decay product \Coa, \Cob and \Ti, are expected to form in 
the deepest layers of the supernova envelope \citep{TimmesTiCo}. Hence, 
their final abundances in the ejected envelope may be depleted with 
respect to the total quantities produced if a significant fraction of 
these inner layers falls back onto the collapsed core while 
the explosion is still progressing. In the simplest picture of black hole 
formation in an otherwise ``successful'' supernova, larger 
remnant masses coincide with more fall back and therefore with lower 
abundances of ejected radioactive isotopes. A quantitative manifestation 
of this assessment is found in survey by \citet{WoosWeav95}. For a wide range 
of parameters, they find that explosions which produce remnants with 
$M_{rem}\ga 3\;M_\sun$ eject negligible abundances of \Ni, while in 
explosions which leave behind remnants of $1-2.5\;M_\sun$, the amount 
of ejected \Ni is of order $0.1\;M_\sun$ (the exact value depends on 
the details of the progenitor and explosion). 
  
\subsubsection{No Radioactive Isotopes}

If we adopt this simple picture of radioactive isotope production in 
supernovae, we may assume that supernovae which produce black 
holes eject negligible amounts of radioactive isotopes. Correspondingly, 
the accretion luminosity must compete only with the thermal emission of the 
internal energy deposited during the explosion. 

Immediately after the explosion 
the envelope is highly ionized, and therefore opaque to its own thermal 
photons. The internal energy is emitted slowly through photon diffusion, 
and the expansion of the envelope is roughly adiabatic, with average density 
and average temperature falling off inversely with time 
\citep{ArnettBook, ZCSW}. Eventually, expansion degrades the envelope 
temperature sufficiently to allow its material to recombine. The recombined 
material is practically transparent to the thermal photons, so the envelope 
cools rapidly. The process is so rapid that the photon 
distribution cannot adjust to recombination - rather, a recombination 
``front'' sweeps through the envelope \citep{ArnettBook}, liberating 
most of its thermal energy.

Recombination sets in after a {\it recombination time}
\begin{equation}\label{eq:t_rec}
t_{rec}=t_0({\rm H})\frac{T_0({\rm H})}{T_{rec}({\rm H})}\;,
\end{equation}
where $T_0({\rm H})$ is the initial average temperature of the hydrogen 
envelope (which contains most of the mass and the internal energy), 
$t_0({\rm H})$ is its initial expansion time and 
$T_{rec}({\rm H})$ is a representative temperature at which hydrogen 
recombines. Roughly, the internal energy is emitted over a time 
$\sim t_{rec}$, during which the recombination front sweeps recedes through 
the bulk of the envelope. We therefore assume that the accretion luminosity 
emerges at a time of roughly
\begin{equation}\label{eq:t_BH2t_rec}
t_{BH}\approx 2 t_{rec}\;.
\end{equation}
This estimate is consistent with observed Type II supernovae (SNeII) 
light curves, where the time until the recombination peak is roughly 
equal to the time it lasts until dropping below luminosity from \Coa emission. 
For example, in SN1987A the recombination peak dominated the light curve 
between about 50 and 120 days after the explosion; in SN1997D, it appears that 
the recombination peak lasted between about 40 and 80 days after the 
explosion. This estimate is also consistent with numerical 
simulations of SNeII light curves, and, specifically, with 
our previous simulations of black hole emergence in the absence of radioactive 
heating \citep{ZCSW, BHinSN}.
     
In the context of our analytic model, we may estimate the recombination 
time by approximating the entire 
ejected material (both the helium-rich layer and the hydrogen-rich envelope) 
as uniform both in temperature and in expansion time, so 
$T_0(H)=T_0,\;t_0(H)=t_0$. Both quantities can then be expressed 
in terms of the initial properties of the progenitor, namely the mass of the 
hydrogen envelope, $M_H$, and the initial outer radius, $R_0$:
\begin{equation}\label{eq:T_0}
E_{th}=\frac{4\pi}{3}R_0^3 aT_0^4,\;\rightarrow\; 
T_0=\left[\frac{3}{4\pi}\frac{E_{th}}{a R_0^3}\right]^{1/4}\;,
\end{equation}
and 
\begin{equation}\label{eq:t_0H}
t_0=\frac{R_0}{V_0({\rm H})}=
R_0 \left[\frac{E_{kin}({\rm H})}{\frac{3}{10}M_{\rm H}}\right]^{-1/2}\;.
\end{equation}  
In equations (\ref{eq:T_0}-\ref{eq:t_0H}), $E_{th}$ is the total initial 
thermal energy in the envelope, while $V_0({\rm H})$ and $E_{kin}({\rm H})$ 
are the outer velocity and the total kinetic energy of the hydrogen-rich 
envelope, respectively. Now $E_{kin}({\rm H})$ is defined by our 
previous parameterization,  
\begin{eqnarray}\label{eq:E_kinH}
E_{kin}({\rm H}) = & \hspace{-2.5cm} E_{kin}-E_{kin}({\rm He}) = \\
 & (1-f_{\rm He})E_{kin}=(1-f_{\rm He})(1-f)E_{exp}  \;. \nonumber
\end{eqnarray}
The thermal energy, $E_{th}$, is some fraction of the 
kinetic energy, $E_{th}=f_{th}E_{kin}$. To be consistent with the 
assumption of homologous ballistic expansion, the thermal energy must be 
significantly lower than the kinetic energy, i.e., $f_{th}\ll 1$; in the 
following we will use the value $f_{th}=0.1$. Note that this choice does not 
necessarily reflect the actual conditions in the supernova at shock break out, 
when the external radius {\it is} $R_0$, since the envelope will not yet have 
settled into a homologous profile. Rather, it is an effective parameterization 
of the post-shock flow once it settles into homologous expansion 
(after a few expansion times; Arnett 1996).

\subsubsection{Some Radioactive Isotopes}  

Even relatively low abundances of \Coa, \Cob and \Ti 
will lead to the tail of the light curve being dominated by radioactive 
heating. In the case of SN1997D the observed amount of 
$2\times 10^{-3}\;M_\sun$ was found to be sufficient to swamp the accretion 
luminosity for about 1000 days after the explosion \citep{SN1997DLet,BHinSN}. 

It is difficult to generalize the significance of the observed \Coa abundance 
in SN1997D, since there have been only a handful of other supernovae 
where the \Coa (\Ni) abundance was observed to be below $0.01\;M_\sun$ 
\citep{Young98}, with SN1997D being the only case where this abundance 
has actually been determined. In view of the importance of 
radioactive heating to the prospects of observing black hole emergence, we 
consider several models of the relation between progenitor mass and 
abundance of radioactive isotopes.
We quantify these models through the abundance of \Coa, which  
dominates radioactive heating after recombination (\Cob and \Ti are discussed 
below). The black hole emerges in the light curve when 
the accretion luminosity equals the luminosity due to \Coa heating, 
$L_{acc}(t_{BH})=L_{\rm ^{56}Co}(t_{BH})$.

As an extreme model, we examine the possibility that due to mixing in 
the earliest stages of the explosion a minute amount of radioactive 
isotopes always reaches the layers which are eventually ejected, regardless 
of the final black hole mass. Hence we assume that all black hole forming 
supernovae eject $M_{\rm ^{56}Co}=2\times 10^{-3}\;M_\sun$, as was inferred 
for SN1997D. As a compromising alternative, we also consider a model where 
there is always finite \Coa abundance which does depend on the black hole 
mass: for lack of an actual physical model, we stipulate that 
$M_{\rm ^{56}Co}=2\times 10^{-3}\;M_\sun$ for $M_{BH}=3\;M_\odot$, but 
declines rapidly with increasing black hole mass through the relation
\begin{equation}\label{eq:M_Co}
M_{\rm ^{56}Co}(M_{BH})=M_0 
\exp \left[-\left(\frac{M_{BH}}{3\; M_\sun}\right)^\beta\right]\;\;,
\end{equation}
assuming $M_{BH}\geq 3\; M_\sun$.
We consider a ``moderate decline'' of $\beta=2$ and a ``rapid-decline'' 
of $\beta=8$. We chose $M_0=5.44\times 10^{-3}\;M_\sun$, so that 
$M_{\rm ^{56}Co}(M_{BH}=3\;M_\sun)=2\times 10^{-3}\;M_\sun$.

At times greater than a few tens of days (when \Ni decay is complete) the 
luminosity due to heating from \Coa decays can be estimated as 
\citep{Woosleyal89} 
\begin{eqnarray}\label{eq:L_Co}
L_{\rm ^{56}Co}(t) & = &\left(\frac{M_{\rm ^{56}Co}}{M_\odot}\right)\times \\
&  & \left[
\varepsilon_\gamma({\rm ^{56}Co})f_\gamma(t)+\varepsilon_{e^+}
({\rm ^{56}Co})\right]\mathrm{e}^{-t/\tau({\rm ^{56}Co})}\;,  \nonumber
\end{eqnarray}
where $\varepsilon_\gamma({\rm ^{56}Co})=1.27\times 10^{43}\;$\ergss and 
$\varepsilon_{e^+}({\rm ^{56}Co})=4.45\times 10^{41}\;$\ergss are the energy 
emission rate in $\gamma-$rays and positrons (produced per $1\;M_\sun$ of 
\Coa), respectively, and 
$\tau({\rm ^{56}Co})=111.3\;$days is the life time of \Coa. 
The coefficient $f_\gamma$ includes the effects of a finite 
optical depth of the envelope to the emitted $\gamma-$energy photons. 
Photons which escape before thermalizing their energy do not 
contribute to the bolometric light curve, i.e., $f_\gamma<1$. Roughly, 
$f_\gamma$ can be estimated through the time dependent optical depth 
of a layer with width $R(t)$:
\begin{equation}\label{eq:f_gamma}
1-f_\gamma(t) \approx \exp [-\!\!\int\limits_{}^{R(t)} 
\kappa_\gamma \rho(r) d\!r ]\;,
\end{equation}
where $\kappa_\gamma$ is the opacity of the material to the \Coa decay 
photons; a good estimate is that for the typical energy of these photons,  
$\kappa_\gamma=0.03(1+X)\;\mbox{cm}^2\;\mbox{gm}^{-1}$, where $X$ is 
the hydrogen mass fraction of the material. 
For a uniform density layer with a mass $M^\prime$ and kinetic energy 
$E_{kin}^\prime$ the $\gamma-$energy photon optical depth 
can be estimated as \citep{BHinSN}:
\begin{equation}\label{eq:optigam}
\int\limits_{}^{R(t)} \kappa_\gamma \rho(r) d\!r  \approx 
\kappa_\gamma R_{out}(t) \rho(t) =
\kappa_\gamma \frac{3}{4\pi}\frac{3}{10}
\frac{{M^\prime}^2}{E_{kin}^\prime} t^{-2}\;,
\end{equation}
and we used $R_{out}(t)=V_0 t$ ($V_0$ is the velocity at the outer edge of 
the layer) and equation (\ref{eq:rho0t03}). 
Note that the relation $f_\gamma\propto 1-\exp(-1/t^2)$ combines with 
the natural exponential to accelerate the 
temporal decline of $L_{\rm ^{56}Co}$ (Eq.~[\ref{eq:L_Co}]), much faster than 
the power-law decline of $L_{acc}$ (Eq.~[\ref{eq:L_acctdust}]).

\subsection{Summary: Model Parameters and Uncertainties}

The parameterizations introduced above allow to determine the 
luminosity at emergence as a function of progenitor mass. The key quantity 
is the effective time $t_{dust}$:
\begin{eqnarray}\label{eq:t_dustfin}
\lefteqn{t_{dust}  =  2.67 \left(\frac{\mu}{0.5}\right)^{-4/5}
       \left(\frac{\kappa}{0.4}\right)^{-3/10}
       \left(\frac{M_{BH}}{M_\sun}\right)^{-1/5}} \\
& &       \left(\frac{M_{\rm He}}{M_\sun}\right)^{5/4} 
       \left(\frac{(1-f)E_{exp}}{10^{51}\;\mbox{ergs}}\right)^{-3/4}
       \left(\frac{f_{\rm He}}{0.01}\right)^{-3/4}\;\;\mbox{days}\;. \nonumber
\end{eqnarray}
In the following we assume that the composition of the bound material 
as a mixture in mass of 0.1 hydrogen, 0.45 helium and 0.45 oxygen, which is 
typical of SNeII models 
(T.~R.~Young, 1999, private communication). For this composition 
$\mu=1.26$ and $\kappa=0.22$.

In the case of no radioactive isotopes in the envelope, the emergence time 
is determined by recombination, and we have
\begin{eqnarray}\label{eq:t_BHrecfin}
\lefteqn{t_{BH} =  2t_{rec}= } \\ \nonumber
& &       138.5\left(\frac{M_{\rm H}}{10M_\sun}\right)^{1/2}
          \left(\frac{R_0}{10^{14}\;\mbox{cm}}\right)^{1/4} 
          \left(\frac{f_{th}}{0.1}\right)^{1/4}           
          \left(\frac{T_{rec}({\rm H})}{10^4\;^{\circ}\mathrm{K}}\right)^{-1}
\\ 
& & \hspace{2.0cm}          
          \left(\frac{(1-f)E_{exp}}{10^{51}\;\mbox{ergs}}\right)^{-1/4}
          \left(1-f_{\rm He}\right)^{-1/4}\;\;\mbox{days}\;.  \nonumber       
\end{eqnarray}

In the case of a finite \Coa abundance, the luminosity due to radioactive 
heating is estimated with equations (\ref{eq:M_Co}-\ref{eq:L_Co}).
The $\gamma-$ray optical depth due to the helium-rich layer and the 
hydrogen-rich envelope is evaluated using equations 
(\ref{eq:f_gamma}-\ref{eq:optigam}). For SNeII we assume that in the 
helium-rich layer $X=0.1$ and that in hydrogen-rich envelope $X=0.7$, giving 
rise to the following result for the $\gamma-$ray trapping factor: 
\begin{eqnarray}\label{eq:f_gammafin}
\lefteqn{\ln(1-f_\gamma(t))=  
     -\left(\frac{t}{300\;\mbox{days}}\right)^{-2}\times}  \\ \nonumber 
& &   \left[1.41\left(\frac{M_{\rm He}}{M_\sun}\right)^2
      \left(\frac{(1-f)E_{exp}}{10^{51}\;\mbox{ergs}}\right)^{-1} 
      \left(\frac{f_{\rm He}}{0.01}\right)^{-1}+\right.\\ 
& & \;\left.2.18\left(\frac{M_{\rm H}}{10M_\sun}\right)^2
      \left(\frac{(1-f)E_{exp}}{10^{51}\;\mbox{ergs}}\right)^{-1} 
      \left(1-f_{\rm He}\right)^{-1}\right]\;\;. \nonumber 
\end{eqnarray}
For most of realistic values of SNeII, the 
contribution of the helium-rich layer (first term in 
eq.~[\ref{eq:f_gammafin}]) dominates $\gamma-$ray trapping, since 
$M_{He}/M_\odot>M_H/10M_\odot$. However, in the low energy 
explosions of the most massive progenitors the contribution of the 
hydrogen-rich envelope becomes comparable, and must be included. 

\subsection{Comments on the Model}

The analytic form of equations (\ref{eq:t_dustfin}-\ref{eq:f_gammafin}) 
enables us to assess the time of emergence and the corresponding luminosity 
for a well defined set of assumptions. Naturally, 
this convenience is achieved at the price of several significant 
simplifications, and we note some obvious sources of uncertainties.
 
{\it Homologous Expansion.} 
The assumption of a homologous expansion 
profile is in itself a crude approximation (see the comparison of an 
analytic estimate for the emergence time and luminosity with the results of a 
simulation with a nonhomologous initial profile in Balberg et al.~2000).
Furthermore, supernova simulations (see, e.g., Woosley 1988, 
Shigiyama \& Nomoto 1990) also suggest that the helium-rich layer and the 
hydrogen-rich envelope do not share a common expansion time, but rather 
$t_0(H)\la t_0(He)$. The ratio of the two tends to be very model 
dependent. Our model must therefore be treated as illustrative, rather than 
exact. 

{\it Secondary Model Simplifications.} 
The lower mass limit for black holes may be significantly less 
than $3\;M_\sun$ \citep{BetheBrown95}. In principle this could add many more 
events, but since lower mass remnants (regardless of nature) are likely to be 
correlated with larger amounts of radioactive isotopes in the ejecta, they 
would not make favorable candidates for observing black emergence. 
Also, there is a small inconsistency since the bound material for late-time 
accretion is actually included in $M_{BH}$ in the recipe of equation 
(\ref{eq:M_rem}). However, since $M_{BH}\geq3\;M_\sun$ while the amount of 
bound material for late-time fallback is $0.1-0.2\;M_\sun$, the impact of this 
inconsistency is small. Finally, we note that 
the estimate of $t_{rec}$ in equation (\ref{eq:t_BHrecfin})
includes two additional parameters, namely the effective thermal energy 
fraction and the progenitor radius, making the 
derivation less general. However, since the dependence on both is weak, 
we do not introduce significant errors by using only 
representative values for these parameters.

{\it Explosion Energy.}
Our one parameter model of $L_{BH}(M_*)$ is obviously a simplification, since 
in reality there should be a range of explosion energies for any 
progenitor mass. Such a range would be due to rotation, metallicity and other 
factors which affect progenitor structure and evolution. We aim to 
accommodate this simplification by conducting a parameter survey with the 
values of $f$ and $f_{He}$, which 
should span a reasonable range of uncertainty in this regard.

{\it Mass Loss.} The model does not include potential mass loss, which is 
expected for massive stars. Mass loss may be important in determining 
the eventual light curve of the supernova, and perhaps even the properties 
of the remnant. 
In this context we distinguish between modest mass loss expected in single 
stars and the substantial mass loss that occurs in a binary system. In the 
former the star is likely to retain a sizable fraction of its hydrogen 
envelope, and so explode as SNII \citep{WillLang99} without a 
significant effect on the explosion energy and the black hole mass 
\citep{MacFetal99,FryKal99}. The more efficient mass loss from massive stars 
in binaries is actually likely to strip the hydrogen envelope from the 
progenitors \citep{WillLang99}, and if it occurs prior to helium ignition, it 
may eventually even affect the nature of the explosion and remnant 
\citep{FryKal99}, complicating a simple estimate along the lines of our model. 
However, once the hydrogen envelope is stripped from the star, the likely 
result will be a Type Ib/Ic supernova, which is unfavorable for detecting 
black hole emergence (see below).
Hence, we can avoid considering the effect of significant mass loss on 
explosion properties and results. On the other hand, modest mass loss does 
increase the likelihood that circumstellar interaction and dust formation 
will obscure the emergence luminosity, and therefore may have an unfavorable 
effect on the prospect of detecting emergence. We return to this point later 
in \S~\ref{Sect:RATES}.

{\it Spherical Accretion.} We must caution 
that the results reported here apply strictly for spherical accretion. 
The assumption of spherical accretion is not necessarily a bad one, since 
late-time accretion settles after many sound-crossing times, and the flow can 
recover from modest asymmetries arising in the explosion. Furthermore, unlike 
accretion in a binary system, the accreting material does not acquire angular 
momentum from inspiral, and the specific angular momentum is limited to 
pre-explosion values, which may not be too large. On the other hand, 
late-time accretion is shaped in part by 
the luminosity at early times, for which disk-like rather than spherical flow 
may be more suitable \citep{Mineshetal97}. The radiative  
efficiency of hypercritical disk accretion is not well constrained by theory, 
and assessing its implications on the prospects of detecting black hole 
emergence is beyond the scope of this paper.

\section{Model Results: Luminosity at Emergence} \label{Sect:LvsM}

With the model of \S~\ref{Sect:Model} we proceed to calculate the 
time of emergence and corresponding luminosity as a function of progenitor 
mass, $t_{BH}(M_*)$ and $L_{acc}(t_{BH})\equiv L_{BH}(M_*)$, 
respectively. 

\subsection{Emergence Luminosity with no Radioactive Isotopes} 
\label{subsect:LBHclean}

The most favorable situation for observing black hole emergence, 
is when radioactive heating is negligible and emergence occurs once 
envelope recombination stratifies. The results for the accretion luminosity 
(Eq.~[\ref{eq:L_acctdust}]) based on the effective time $t_{dust}$ 
(Eq.~[\ref{eq:t_dustfin}]) and emergence time, $t_{BH}$ 
(Eq.~[\ref{eq:t_BHrecfin}]) as a function of progenitor mass are shown in 
Figure~\ref{fig:LvsMclean}. In the calculations we varied the fraction of 
explosion energy spent on unbinding the star ($f=0.25,\;0.50$ and 0.75), 
and the fraction of kinetic energy deposited in the helium layer  
($f_{\rm He}=0.01$ and $0.03$). We set the progenitor radius at 
$R_0=5\times 10^{13}\;$cm and the fiducial fraction of thermal energy as 
$f_{th}=0.1$.

The accretion luminosities at emergence varies between several 
$10^{35}$ to $3\times10^{37}\;$\ergss for the most optimistic models. The 
results are quite sensitive to both effective parameters $f$ and $f_{\rm He}$, 
and, for a given progenitor mass, can vary by more than an order of magnitude 
between extreme assumptions. In spite of these uncertainties, we 
can identify some global trends. First, for a given total explosion 
energy and black hole mass (given value of $f$), the luminosity at emergence 
increases with decreasing $f_{\rm He}$: less energy to the helium layer means 
more bound mass which is also slowly moving as a reservoir for late-time 
accretion. Within the parameterization of the model, 
$t_{dust}\propto f_{\rm He}^{-3/4}$ while $t_{BH}$ is very weakly dependent on 
$f_{\rm He}$, thereby yielding the relation 
$L_{BH}\propto f_{\rm He}^{75/72}$, which holds for progenitors of all masses. 

The luminosity at emergence also increases with increasing $f$, 
and that the sensitivity to this factor is at least as large as that to 
$f_{\rm He}$. Although a larger available energy for 
unbinding the star leads to a lower mass black hole, it implies a larger 
helium-rich layer in the ejecta and less kinetic energy for the ejecta 
(which scales as $1-f$). In the parameterization 
equations~(\ref{eq:L_acctdust}) and (\ref{eq:t_dustfin}-\ref{eq:t_BHrecfin}) 
a thick and slow helium-rich layer (i.e., a larger reservoir of material for 
late-time accretion) outweighs
\begin{inlinefigure}
\centerline{\includegraphics[width=1.0\linewidth]{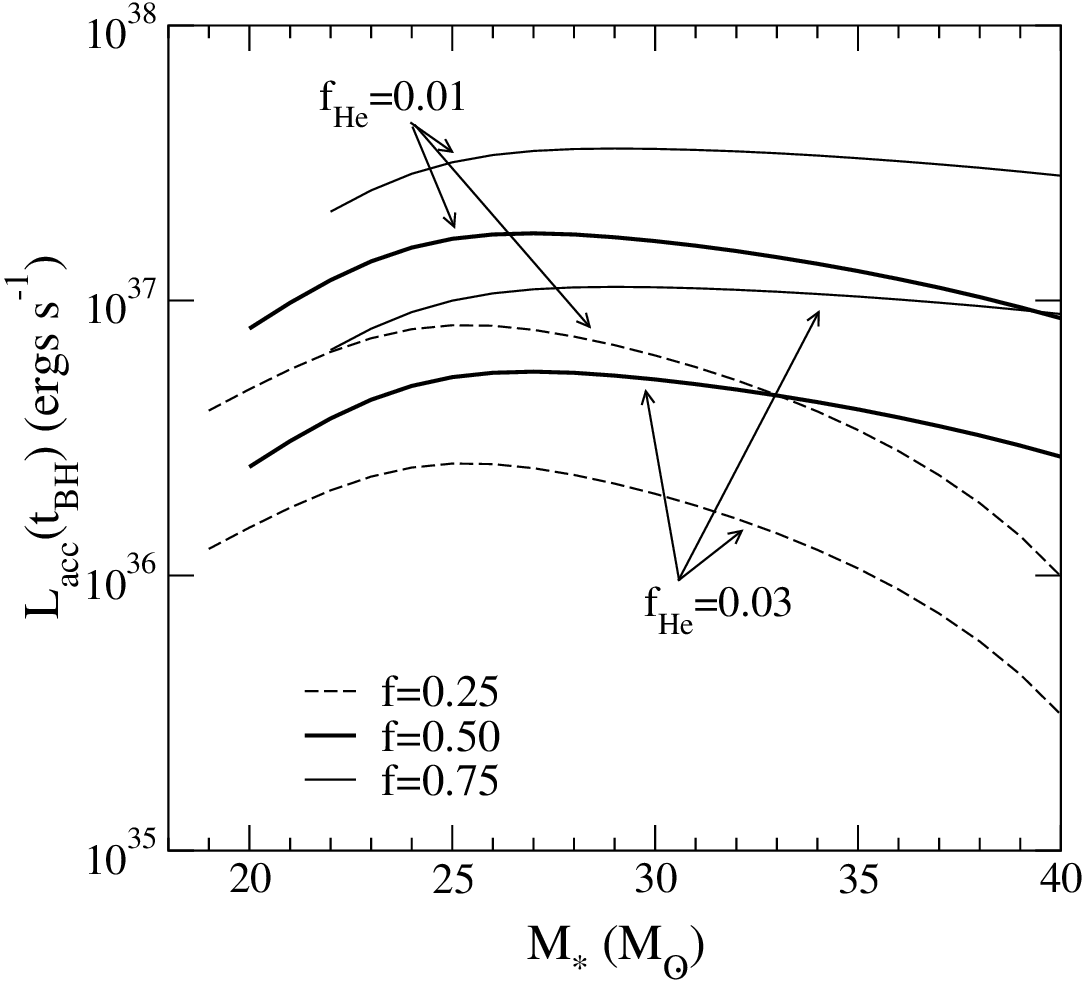}}
\figcaption{
Parameter survey for the accretion luminosity at emergence, $L_{BH}$, as a 
function of progenitor mass, $M_*$, for the model with no radioactive heating. 
Parameters surveyed are $f$ and $f_{\rm He}$, with $f=0.25,\;-$ dashed lines, 
$f=0.50\;-$ thick solid lines, and $f=0.75\;-$ thin 
solid lines, and with $f_{\rm He}=0.01$, and 0.03.\label{fig:LvsMclean} }
\end{inlinefigure}

\setlength{\parindent}{0.0in}
the mass of the black hole in determining 
the magnitude of the accretion luminosity. For realistic values of $f$ 
we find that roughly $L_{BH}\propto f$.
This result reinforces previous expectations 
\citep{ZCSW,BHinSN} that supernovae with low inferred explosion energies 
(low expansion velocities) are the most promising in terms of observing black 
hole emergence. Note that since the actual values of black hole and 
helium-layer masses also depend on the progenitor mass, changing $f$ does 
not just scale the $L_{BH}(M_*)$ curve (as changing $f_{\rm He}$ does) but 
also modifies the shape of this function. The dependence of emergence 
luminosity on progenitor mass can also be understood in terms of the 
competition between black hole mass and ejected helium layer mass.  
\parindent=3.5mm

\subsection{Emergence Luminosity with Radioactive Heating}
\label{subsect:LBHrad}

\begin{figure*}
\vspace{-0.25cm}
\plotone{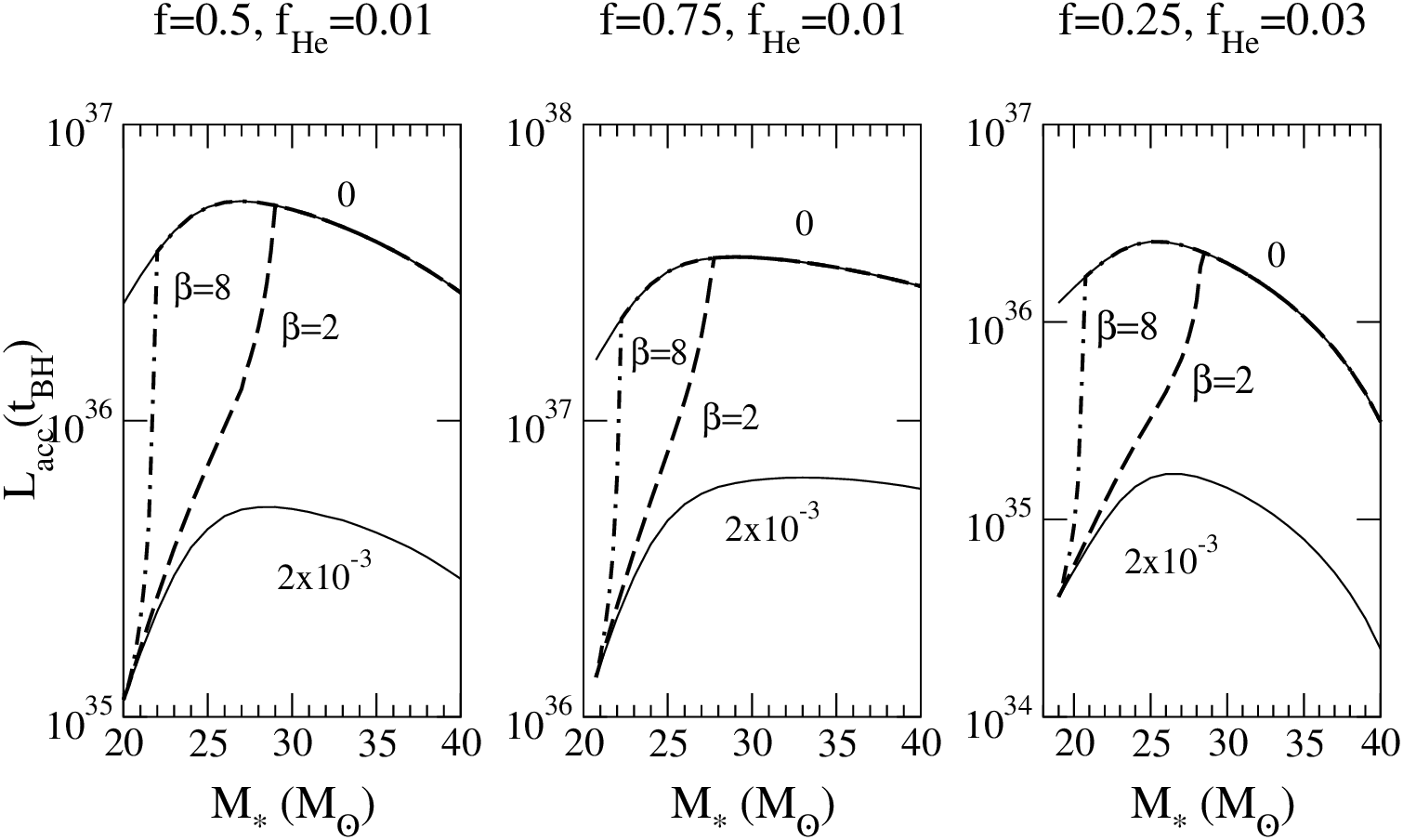}
\caption{
Accretion luminosity at emergence, $L_{BH}$, as a function of progenitor mass 
$M_*$ for three combinations of $f$ and $f_{\rm He}$ and different assumptions 
about the abundance of \Coa in the ejected envelope. Curves are labeled with 
the mass of $M_{\rm ^{56}Co}$ if independent on $M_*$ and by the value of 
$\beta$ used for a $M_{\rm ^{56}Co}(M_*)$ according to its value in 
equation (\ref{eq:M_Co}).\label{fig:LvsMRad}}
\vspace{-0.25cm}
\end{figure*}

If radioactive decays dominate over accretion as the heating source 
after recombination, emergence can be significantly delayed.
We determine emergence time and luminosity by comparing the evolution of 
the accretion luminosity (eqs.~[\ref{eq:L_acctdust}] and [\ref{eq:t_dustfin}]) 
and the radioactive luminosity (eqs.~[\ref{eq:L_Co}] and 
[\ref{eq:f_gammafin}]) for a given $M_{\rm ^{56}Co}$ (eq.~[\ref{eq:M_Co}]) in 
the ejecta. The emergence luminosity as function of progenitor mass, 
$L_{BH}(M_*)$, is shown for three combinations of $f$ and $f_{\rm He}$ in 
Figure~\ref{fig:LvsMRad}; these are $\{f=0.5\;f_{\rm He}=0.01\}$, 
$\{f=0.75\;f_{\rm He}=0.01\}$ and $\{f=0.25\;f_{\rm He}=0.03\}$, which are 
intermediate, most optimistic and least optimistic, respectively, in terms of 
the accretion luminosity they predict. In each case we compare four 
assumptions about the abundance of \Coa: none (i.e., the results in 
\S~\ref{subsect:LBHclean}), $M_{\rm ^{56}Co}=2\times 10^{-3}\;M_\odot$ for all 
progenitors, and a black hole mass-dependent abundance according to 
equation~(\ref{eq:M_Co}), with $\beta=2\;$and 8 (labeled with the value of 
$\beta$). 

The dominant feature of Figure~\ref{fig:LvsMRad} is that the qualitative 
effect of including radioactive heating is similar for all three cases (and, 
indeed, is similar for other combinations of $f$ and $f_{\rm He}$ not shown 
here). When a constant $M_{\rm ^{56}Co}=2\times 10^{-3}\;M_\odot$ is assumed 
for all progenitor masses, the emergence luminosity is decreased by a factor 
of 5-50 with respect to its value in the radioactive-free case for the same 
$\{f,\;f_{\rm He}\}$ combination; a greater decrease corresponds to a lower 
value of $L_{BH}$ in 
the radioactive-free case. This similarity results 
because the time of emergence is controlled by the exponential 
decline of the \Coa heating rate; the emergence times are then constrained to 
the rather narrow range of $1200\pm 200$ days, hence constraining the 
value of $L_{BH}$ as well.

When the amount of \Coa is assumed to decline with increasing black hole mass, 
the $L_{BH}$ curve ascends from the $M_{^{56}Co}=2\times 10^{-3}\;M_\odot$ 
values for lower mass progenitors to the radioactive-free values for higher 
mass progenitors; the slope of this ascent is, of course, determined in our 
parameterization by the value of $\beta$. The net effect of such a black hole 
mass-dependent \Coa abundance is to suppress the contribution of 
the lower mass range of progenitors to observable black hole emergence events. 
The implications of this suppression on the rate of observable 
events of black hole emergence may be quite significant, if a realistic 
progenitor mass distribution is weighted towards the lower mass stars, these 
progenitors dominate the supernova rate. 

It is important to comment on the implications of \Cob and \Ti. 
If those isotopes are present with abundances which scale with 
$M_{\rm {^56}Co}$ similar to the ratios inferred in SN1987A \citep{WoosTim96}, 
they would dramatically alter the prospects for observing black hole 
emergence. It is unclear to what extent such scaling is justified, since \Cob 
and \Ti are synthesized even deeper than \Ni, 
so scaling is mostly useful for placing an upper limit on the \Cob and \Ti 
abundances. If these abundances do scale according to
$M_{\rm ^{56}Co}=2\times 10^{-3}\;M_\odot$ and the SN1987A ratios, 
both become important for radioactive heating at about 1000 days 
after the explosion \citep{BHinSN}, with heating rates of  
$L_{\rm ^{57}Co}\approx L_{\rm ^{44}Ti}\lesssim 10^{35}\;$\ergss. Heating from 
\Ti decay (positron heating) should remain at this level for tens of years 
after the explosion. Thus, if $L_{BH}$ estimated from comparison with 
$L_{\rm ^{56}Co}$ is slightly less than a several $10^{35}\;$\ergss, unveiling 
the accretion luminosity would be a more difficult task since the total 
luminosity will include additional components (rather 
than follow a simple power law decay). If $L_{BH}$ is significantly less than 
$10^{35}$\ergss, as in the case of $f=0.25\;f_{\rm He}=0.03$ and   
$M_{\rm ^{56}Co}=2\times 10^{-3}\;M_\odot$, the accretion luminosity would be 
sufficiently subdominant so that detection would be impossible. Fortunately, 
we can avoid detailed investigation of the effects of \Cob and \Ti since 
for HST capabilities an emergence luminosity below $10^{35}\;$\ergss cannot 
be detected beyond the Local Group (see below). 
Provided that scaling of \Cob and \Ti abundances as in SN1987A applies, 
any emergence luminosity that is low enough to be obscured by \Cob and \Ti 
heating is undetectable in any case.

\section{Rate of Observable Black Hole Emergence}
\label{Sect:RATES}

Having derived the function $L_{BH}(M_*)$, we now proceed to estimate the 
event rate. We define an ``event'' in this context as 
a case where the accretion luminosity at emergence is above the threshold of 
a given instrument (the total luminosity is twice as large, of course, but we 
assume conservatively that a factor of two margin would be needed to certify 
the nature of the accretion luminosity, especially if the power-law 
dependence is to
\begin{figure*}
\vspace{-0.25cm}
\plotone{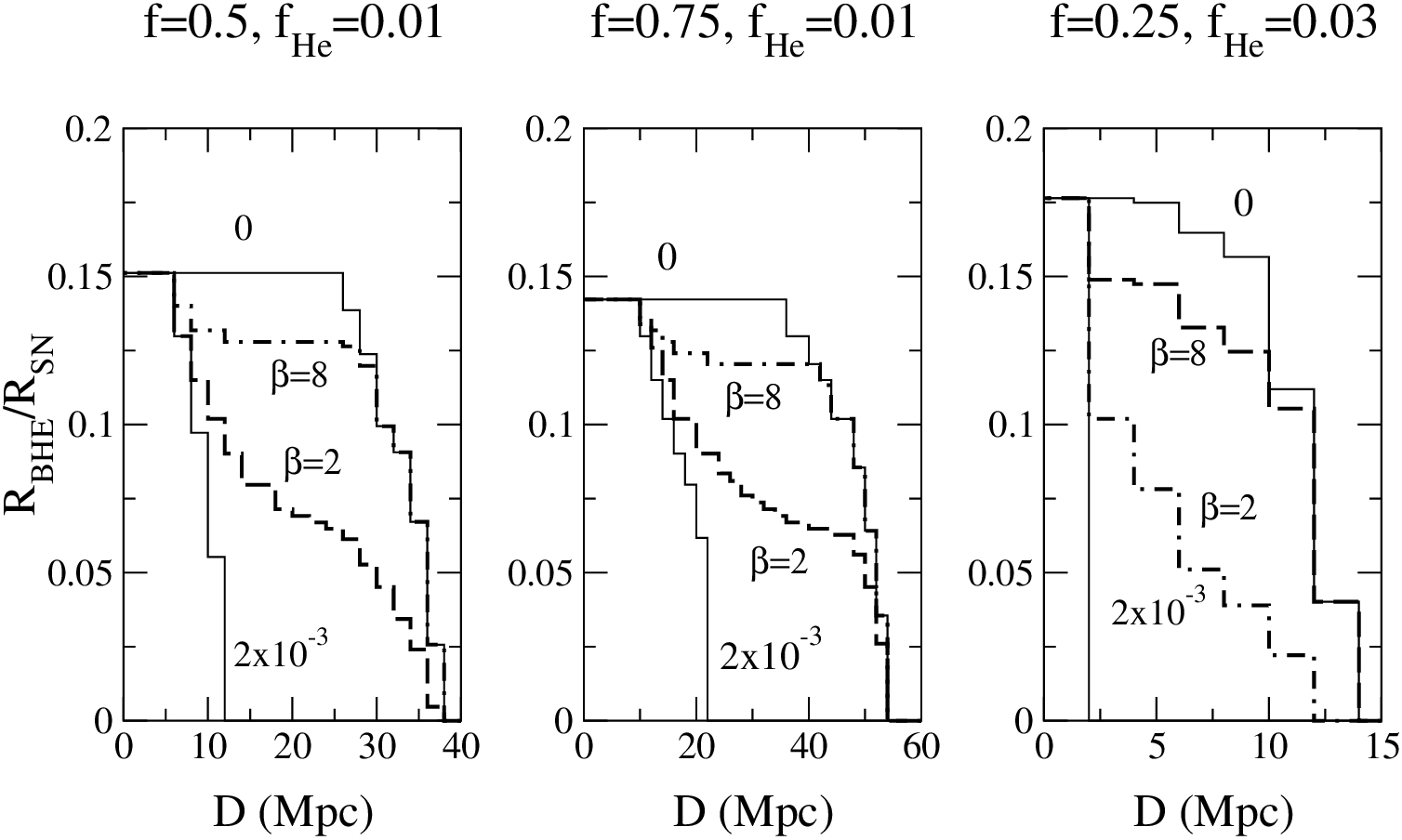}
\caption{
The fraction of Type II supernovae for which black hole emergence 
can be observed as a out of the all Type II supernovae as a function 
of distance for three combinations of $f$ and $f_{\rm He}$ and the 
various assumptions about the \Coa abundance (see Fig.~\ref{fig:LvsMRad}) 
and $\gamma=2.7$.\label{fig:fracdis}}
\vspace{-0.25cm}
\end{figure*}
be identified). The detector 
threshold translates into a minimum emergence luminosity which can be detected 
from any given distance. We will use the HST STIS camera capabilities as a our 
reference and require a threshold apparent bolometric magnitude of $m_B=28.5$. 
The function $L_{BH}(M_*)$ provides the progenitor masses which have emergence 
luminosity exceeding that minimum luminosity; their relative fraction 
out of all progenitors which make core collapse supernovae is determined by 
assuming a present day mass function (PDMF) of progenitors. 
In this section we estimate this fraction and compare it with the observed 
supernova rate in the local universe in completing an estimate 
of events of observable black hole emergence. 

\vspace{0.30cm}
\subsection{Events per Supernova as a Function of Distance}
\label{subsect:fracDIS}

For simplicity, we assume a single power law PDMF for the progenitors of 
core collapse supernova, similar to the initial mass function (IMF) of the 
Milky Way. There is some observational evidence the 
mass function of massive stars in the Milky Way does follow such a power law 
with no sign of dependence on metallicity \citep{MasThom91}.
In this parameterization, the fraction $F(M)$ of 
progenitors with mass greater than $M$ follows the relation 
\begin{eqnarray}\label{eq:PDMF}
{\frac{dF}{dM}\propto M^{-\gamma}\;;  
F(M)=C_{cc}M^{-\gamma+1}\;;}   \\
{C_{cc}=(M_1^{-\gamma+1}-M_2^{-\gamma+1})^{-1}\;,} \nonumber
\end{eqnarray}
where the value of the constant $C_{cc}$ arises from stipulating that 
core-collapse supernovae occur in stars with progenitor masses between $M_1$ 
and $M_2$.

We also assume that all local galaxies which 
host core-collapse supernovae share a single power law $\gamma$. By changing 
$\gamma$ over a reasonable range of values we can assess the likely 
uncertainties embedded in this assumption. The value of $\gamma$ is crucial 
in determining the black-hole-per-core-collapse-supernova fraction, which is 
essentially the fraction of progenitors which leave behind a remnant with 
$M\geq3\;M_\odot$. By following \citet{FryKal99} who adopted the 
\citet{Scalo} value of $\gamma=2.7$ as a nominal choice and assumed core 
collapse supernova occur for progenitors with masses in the range 
$8-40\;M_\odot$, we find that the fraction of supernovae which make black 
holes is about 0.15. 
 
The combination of the function $L_{BH}(M_*)$ and the PDMF of 
progenitors allows to estimate what is the fraction of black hole forming 
supernovae for which emergence is observable as a function of 
distance. It can be further expanded to estimate 
the fraction of emergence events out of all core-collapse supernovae (most of 
which produce neutron stars). In Figure~\ref{fig:fracdis} we show this 
fraction as a function of distance from the Milky Way, (for a 
threshold apparent magnitude of $28.5$), assuming no dust 
extinction along the line of site.
The three graphs correspond to the three combinations of $f$ and 
$f_{\rm He}$ shown in Figure~\ref{fig:LvsMRad}, with each case including the 
different assumptions concerning the abundance of \Coa. The results are 
binned in $2\;$Mpc intervals. 

At very close distances, all supernovae which create black holes yield 
an observable emergence, so the ``event'' per supernova rate is simply the 
fraction of supernovae which make black holes. 
The characteristic fraction of $\sim15\%$ for $\gamma=2.7$ does have some 
dependence on the value of $f$, which determines the threshold progenitor mass 
for producing a black hole. As distance is increased, for some progenitor 
masses the emergence luminosity falls below threshold and the event fraction
\begin{figure*}
\vspace{-0.25cm}
\plotone{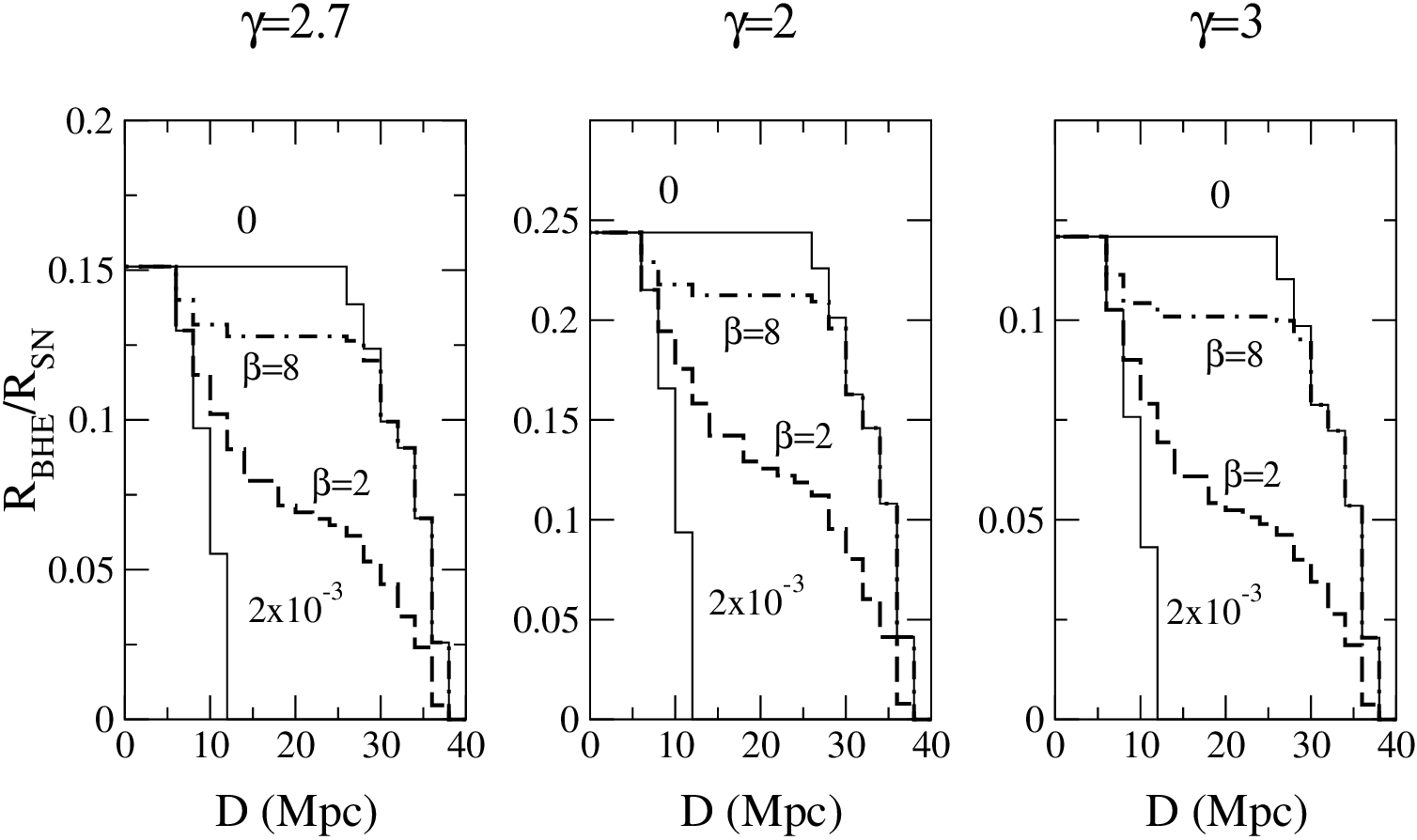}
\caption{ 
The fraction of Type II supernovae for which black hole emergence 
can be observed as a out of the all Type II supernovae as a function of 
distance for for $\gamma=2.7$ (nominal) and 2 and 3, the 
various assumptions about the \Coa abundance (see Fig.~\ref{fig:LvsMRad}), 
and \{$f=0.5\; f_{\rm He}=0.01$\}. \label{fig:disbin_gam}}
\vspace{-0.25cm}
\end{figure*} 
(out of all supernovae) decreases. The significance of radioactive heating is 
clearly evident in the details of each graph, where a larger $M_{\rm ^{56}Co}$ 
naturally delays the emergence of the black hole, and so the luminosity at 
emergence is lower and therefore is observable over smaller distances. The 
importance of the distinction between the two exponential cases, $\beta=2$ 
and $\beta=8$ is also evident. For $\beta=8$ radioactive heating does not 
affect the maximum values of $L_{BH}$ so the dominant contribution to
cases of observable emergence, especially at large distances, is similar 
for the radioactive-free case and the $\beta=8$ case. On the other hand the 
$\beta=2$ model results in reducing the maximum value of $L_{BH}$, thereby 
reducing the maximum distance, $D_{max}$, of observable emergence. 

Most notable is the difference between typical distances over 
which emergence is observable for the different combinations of $f$ and 
$f_{\rm He}$. These vary from $D_{max}\approx 14\;$Mpc for the worst case 
$\{f=0.25\;f_{\rm He}=0.03\}$ (and radioactive free) to 
$D_{max}\approx 54\;$Mpc for the best case $\{f=0.75\;f_{\rm He}=0.01\}$,  
which roughly amounts to a factor of about 60 in volume, and hence in the 
event rate. In general, we find that the combination of 
the uncertainty in $f$ and $f_{\rm He}$ and in the abundance of \Coa places 
$D_{max}$ roughly in the range of $10-50\;$Mpc, corresponding to a 
span in the expected event rate of two orders of magnitude.
In this last assessment we exclude the case of $\{f=0.25\;f_{\rm He}=0.03\}$ 
and $M_{\rm ^{56}Co}=2\times 10^{-3}\;M_\odot$ for which black hole emergence 
is essentially unobservable beyond the Local Group. Recall that this is also 
the combination where the accretion luminosity may be obscured if the amount 
of \Cob and \Ti is not negligible. If this extreme combination is 
representative of black hole forming supernovae, the rate of observable 
emergence is very small.

The slope of the PDMF, $\gamma$, has some influence over the fraction 
of black hole-forming supernovae and over the observable event rate. 
For values of $\gamma$ between 2 and 3, the fraction of core-collapse 
supernovae which form black holes varies between 0.25 and 0.125, 
respectively. The point is that the value of $\gamma$ does not affect the 
maximum distance for observation for a given $\{f\;f_{\rm He}\}$ model. 
Figure~\ref{fig:disbin_gam} compares the event per core collapse supernova 
fraction for $\{f=0.5\;f_{\rm He}=0.01\}$ for $\gamma=2,\;2.7$, and 3. In all 
cases the fraction at any given distance approximately follows the total 
fraction of progenitors which make black holes. Note that the graphs 
for $\gamma=2$ and $\gamma=3$ are almost identical, after the y-axes are 
scaled 2:1.In total, the uncertainty in $\gamma$ accounts 
for a relatively modest uncertainty of a factor of 2 in the event rate.

An additional modest source of uncertainty is the lower limit on 
progenitor masses which make core collapse supernovae. For 
$\gamma=2.7$, stars between 8 and $10\;M_\odot$ make up about one third of 
all those between $8\!-\!40\;M_\odot$, so the fraction of supernovae from 
higher mass progenitors out of the total could be sensitive to our choice 
of the lowest mass progenitors by as much as 50\%.
On the other hand, unless the PDMF is exceptionally flat, the exact value of 
the upper mass limit is not too important, since stars at the very high mass 
end are very rare. 

\subsection{Eliminating Type Ib/c Supernovae}

So far we referred to estimates of the event rate per core-collapse supernova, 
which is the expected fate of massive progenitors. These include 
also Type Ib/c supernovae which are not good candidates for 
observing black hole emergence \citep{BHinSN}: if the hydrogen 
envelope has been ejected prior to explosion, practically all the 
kinetic energy is deposited in the helium layer $(f_{\rm He}=1)$. Effectively, 
the evolutionary option of a Type Ib/c supernova reduces the fraction of 
core collapse supernovae in which black hole emergence could be observable. 

Studies of single and binary stellar evolution indicate 
that the likelihood that sufficient mass loss will occur during the life time 
of a massive star depends mainly on the presence and properties 
of a binary companion; the mass of the star appears to play a secondary role. 
In our analysis we use the simplest classification offered by 
\citet{WillLang99}, where massive stars in binaries lose all their hydrogen 
envelope and result in a Type Ib/Ic supernova, and single massive stars do not 
lose all of the envelope, giving rise to a SNII. This simplest interpretation
is that the probability of a massive star yielding a SNII, $P_{II}$, is 
independent of its mass. With this assumption, the calculations of the
fractions shown in Figures~\ref{fig:fracdis}-\ref{fig:disbin_gam} are then 
the fractions {\it per SNII}, related to the fraction {\it per core-collapse 
supernovae} by a factor of $P_{II}$. In the following we will set 
$P_{II}=0.85$, which is roughly consistent with theoretical estimates of 
supernova rates \citep{Thielemann94, CapEvTur99}.

\subsection{Rate of Events in the Local Universe}

If the efficiency of discovering supernovae in the local universe ($D\approx 
50\;$Mpc) were perfect, event-per-supernova fractions such as those shown in 
Figure~\ref{fig:fracdis} should be 
compared to the estimated supernova rate per galaxy. These are somewhat 
constrained based on observation and theory, and generally expressed in units 
of SNu (1SNu = supernovae per century per $10^{10}\;L_\odot$). 
The estimated rates \citep{Thielemann94, CapEvTur99} are dependent upon galaxy 
type, where core collapse 
supernovae appear to be limited to spiral and irregular galaxies. The combined 
theoretical rates (in SNu) for spiral galaxies are: SNeII - about 
$0.7\pm 0.3$, and Type Ib/c - about $0.14\pm 0.07$ (with some dependence 
on specific galaxy type). In irregular and peculiar galaxies there appears to 
be a slightly larger fraction of Type Ib/c supernovae (SNeIb/Ic), whereas in 
elliptical there are hardly any core collapse supernovae at all. 

However, the actual constraint on observing black hole emergence rate 
is the rate at which supernovae are {\it discovered}, rather than 
the rate at which they occur. Clearly, any attempt to uncover the 
emergence of the black hole would be conditional on the supernova being found 
in the first place. The efficiency of discovering supernovae is imperfect due 
to both incomplete monitoring of the sky and line of sight 
limitations (background, extinction etc.). For example, unfavorable 
inclination of the host galaxy appears to cause a deficiency of observed 
supernova in galactic cores with respect to those found which are offset 
from the galactic centers \citep{Cappellaro97}. 

\subsubsection{Rate of Observed Supernovae}

A recent compilation of supernovae catalogs \citep{SNAsiago99} suggests that 
the recent (1997-1998) rate of {\it observed} supernovae identified as core 
collapse in the local universe is roughly $10^{-4}\;$Mpc$^{-3}\;$yr$^{-1}$ 
although the uncertainty in this estimate is quite large. 
This number includes both SNeII and SNeIb/Ic, with a ratio of 
about 5:1, respectively. Within the statistical uncertainties of the search, 
these rates appears to be independent of distance up to 50Mpc. Such 
distance-independence arises from the limits of supernovae searches 
(incomplete coverage of the sky, methods of the searches), which  
currently dominate over the dimming of the more distant supernovae. 
A rate of $10^{-4}\;$Mpc$^{-3}\;$yr$^{-1}$ is roughly 
10-20\% of the total rate of core collapse supernovae which are expected 
to occur in the local universe.

While we will use this estimate of $10^{-4}\;$Mpc$^{-3}\;$yr$^{-1}$ core 
collapse supernova as a guideline in our calculation, it is clear that at 
very small distances any approximation of a uniform observed supernova rate 
must break down. In particular, the actual supernova rate up to several Mpc 
from the Milky Way is dominated by a few individual galaxies: most notably, 
the star burst galaxies NGC 253 (3.0 Mpc) and  M82 (3.5Mpc). We rely on the 
estimates of \citet{Becklin91}, who evaluated the core-collapse supernova rate 
up to 12 Mpc from the properties of observed galaxies. His estimate was based 
on a rather high theoretical rate of core collapse supernovae: 6 per century 
per $10^{10}\;L_\odot$ in IR (the contribution of massive stars to IR emission 
exceeds that in optical). Recent surveys seem to favor a rate which is about 
one half of that. We therefore modify those estimates by a factor of 0.5. 
The rates of core collapse supernovae that we use for this range are: 
2-4 Mpc - $0.3$yr$^{-1}$; 4-6 Mpc - $0.02$yr$^{-1}$; 
6-8 Mpc - $0.4$yr$^{-1}$; 8-10 Mpc $0.15$yr$^{-1}$; and 10-12 Mpc - 
$0.3$yr$^{-1}$. This last value fits onto the estimated observed rate of 
$10^{-4}\;$Mpc$^{-3}\;$yr$^{-1}$ assumed to hold beyond these closest 
distances. We do not include here supernovae in the Local Group (0-2 Mpc), 
since they are dominated by the Milky Way and for the most part would be 
unobservable. 

\subsubsection{The Rate of Observable Black Hole Emergence Events}
 
We first assume naively that the probability 
of detection of a SNII is independent of progenitor mass (hereafter the 
UNBIASED model).
The event fraction per {\it occurring} core collapse supernova in 
\S~\ref{subsect:fracDIS} then translates simply into an event fraction per 
{\it observed} core collapse supernovae. Using the estimates for the observed 
rate above, we then  calculate the total rate of observable 
black hole emergence events, including a factor of 0.85 to count only SNeII 
where emergence is observable. These estimated rates as a function of 
$f,\;f_{\rm He}$ and the assumptions regarding the \Coa abundance are shown in 
Table~\ref{tab:navRATES}.
\begin{inlinetable}
\caption{Predicted rates (yr$^{-1}$) for observable black hole emergence 
in the ``naive'' model (see text). \label{tab:navRATES}}
\begin{center}
\begin{tabular}{l l | c c c c}
\hline\hline
$f$ & $f_{\rm He}$  
& $M_{\rm ^{56}Co}=$ & $M_{\rm ^{56}Co}=$ & $\beta=2$ & $\beta=8$  
\\  &
&         0          & $2\times 10^{-3}\;M_\odot$
& (Eq.~[\ref{eq:M_Co}])
& (Eq.~[\ref{eq:M_Co}])\\
\hline
0.25 & 0.01 & 0.72 & 0.06     & 0.28 & 0.64  \\
0.25 & 0.03 & 0.23 & $\sim 0$ & 0.08 & 0.19  \\
0.50 & 0.01 & 2.05 & 0.17     & 0.97 & 1.89  \\
0.50 & 0.03 & 0.43 & 0.05     & 0.19 & 0.37  \\
0.75 & 0.01 & 5.74 & 0.44     & 3.45 & 5.36  \\
0.75 & 0.03 & 1.06 & 0.13     & 0.58 & 0.97  \\
\hline\hline
\end{tabular}
\end{center}
\end{inlinetable}

The results in Table~\ref{tab:navRATES} indicate that 
the likely rate of detecting of black hole emergence in 
supernovae in nearby galaxies clearly depends on all three of our main 
unknowns - the fraction of energy spent on unbinding the stars, distribution 
of kinetic energy between the hydrogen envelope and the helium layer, and 
the abundance of \Coa. Within the limits we used in our models, 
each parameter alone imposes an uncertainty of a factor of a few 
in the event rate, whereas the combined effect of all three accumulates to 
about two orders of magnitude of uncertainty. The prediction for the event 
rate varies between as many 
\begin{figure*}
\vspace{-0.25cm}
\plotone{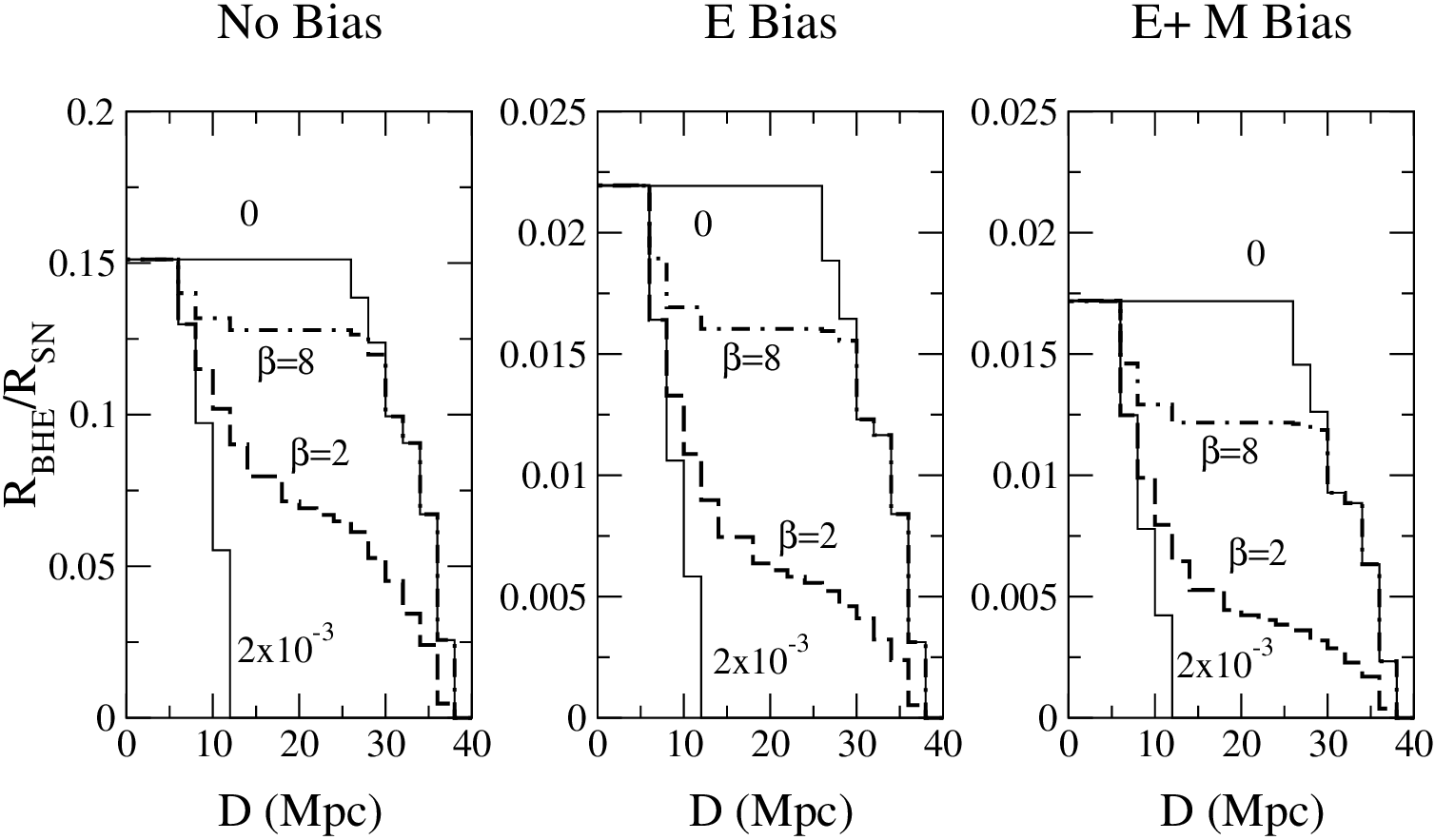}
\caption{
The fraction of {\it observable} Type II supernovae for which black hole 
emergence can be observed as a function of distance out the total number of 
{\it observable} Type II supernovae for three sets of assumptions about 
supernova observability and accretion luminosity. These are  (from 
left to right) - UNBIASED: (same as left panel in Fig.~\ref{fig:fracdis}); 
E-BIAS: underdetection of low energy explosions, and E+M-BIAS: underdetection 
of low energy explosions combined with CSM obstruction of black hole emergence 
(see text). In all cases $f=0.5$, $f_{\rm He}=0.01$ and $\gamma=2.7$; all 
variations for the \Coa abundance are shown (see Fig.~\ref{fig:LvsMRad}). 
\label{fig:DISBINBi}}
\vspace{-0.25cm}
\end{figure*}
as a few events per year to a few events per 
century. In the former case, one expects that several of the SNeII 
discovered each year will lead to an observable black hole 
emergence, were they to be monitored for months after the explosion. Note 
that for most of the parameter range, the event rate should at least be one 
event every few years; rates lower than that are confined almost uniquely 
to the worst case model for radioactive heating, when a constant 
$M_{\rm ^{56}Co}=2 \times 10^{-3}\;M_\odot$ for all supernovae is assumed. 
 
\subsubsection{Effects of Bias in Observability of Supernovae and Emergence}

The estimates above may be too naive in their prediction 
of the rate of observable events of black hole emergence, since it 
explicitly assumes that all SNeII are equally likely to be 
detected. A more realistic point of view is that the 
probability of detecting SNeII is not uniform, but rather is 
dependent on its peak luminosity. Recall that the best 
known candidate for observing black hole emergence, SN1997D, is considered 
significantly subluminous in comparison with typical SNeII. 
While it is unlikely that underluminous supernova 
dominate the total supernova rate (Cappellaro et al.~1997 limit them  
at $20-30\%$ of the total rate), they may represent a large fraction of the 
black hole forming explosions. If, on average, black hole forming supernovae 
have a lower probability of being discovered, the fraction of events per 
observed SNII is obviously reduced with respect to our first estimate.

The luminosity of supernovae during the diffusion and recombination phase is 
known to depend on the explosion energy and progenitor radius and structure 
\citep{ArnettBook}, and indeed, observations indicate a heterogeneous 
luminosity distribution. In particular, relatively compact progenitors will 
result in an underluminous light curve even if the explosion energy was 
quite large, as was the case in the very well studied SN1987A. Nonetheless, 
it is reasonable to expect some correlation in the detectability of supernovae 
with explosion energy. For lack of a better model, we assume that the 
probability of detecting the supernova is proportional its typical luminosity. 
We further assume that 
this typical luminosity is proportional to explosion energy deposited in the 
envelope (i.e., to $(1-f)E_{exp}$ in our parameterization). This approximation 
is roughly correct for a given progenitor \citep{ArnettBook,BHinSN}. 
We show the 
quantitative effect of such a bias against detecting low luminosity 
supernovae on the resulting fractions of event per observed SNII as a function 
of distance in Figure~\ref{fig:DISBINBi}b, for $\gamma=2.7$, 
$\{f=0.50\;f_{\rm He}=0.01\}$ and after normalizing the fraction of all 
observed supernovae in the range $8-40\;M_\odot$ to unity. The immediate 
result in this model (hereafter the E-BIAS model) is a significant 
reduction of the typical fraction of event per observed supernova, which is 
now down to about $0.02$ instead of $0.15$. Also, the relative impact of the 
bias is larger in the cases with a progenitor mass dependence of the \Coa 
abundance. This is quite expected, since in these cases the effect of \Coa is 
limited to the lower mass progenitors, so the relative contribution of the 
higher mass progenitors in the event is greater. 
The corresponding predicted rates under this E-BIAS model 
for the various $\{f\;f_{\rm He}\}$ combinations are shown in 
Table~\ref{tab:BiERATES}.
\begin{inlinetable}
\vspace{0.25cm}
\def\arraystretch{0.8}%
\setlength{\tabcolsep}{5pt}
\caption{Predicted rates (yr$^{-1}$) for observable black hole emergence 
in the model with an explosion energy Bias (see text - model E-BIAS). 
\label{tab:BiERATES}}
\begin{tabular}{l l | c c c c}
\hline\hline
$f$ & $f_{\rm He}$  
& $M_{\rm ^{56}Co}=$ & $M_{\rm ^{56}Co}=$ & $\beta=2$ & $\beta=8$  
\\  &
&         0          & $2\times 10^{-3}\;M_\odot$
& (Eq.~[\ref{eq:M_Co}])
& (Eq.~[\ref{eq:M_Co}])\\
\hline
0.25 & 0.01 & $1.51\!\times\!10^{-1}$ & $9.83\!\times\!10^{-3}$ & $4.02\!\times\!10^{-2}$ & $1.24\!\times\!10^{-1}$  \\
0.25 & 0.03 & $4.91\!\times\!10^{-2}$ &     $\sim 0$        & $9.90\!\times\!10^{-3}$ & $3.65\!\times\!10^{-2}$  \\
0.50 & 0.01 & $2.80\!\times\!10^{-1}$ & $2.22\!\times\!10^{-2}$ & $9.56\!\times\!10^{-2}$ & $2.40\!\times\!10^{-1}$  \\
0.50 & 0.03 & $5.90\!\times\!10^{-2}$ & $5.08\!\times\!10^{-3}$ & $1.78\!\times\!10^{-2}$ & $4.69\!\times\!10^{-2}$  \\
0.75 & 0.01 & $3.76\!\times\!10^{-1}$ & $9.83\!\times\!10^{-2}$ & $1.72\!\times\!10^{-1}$ & $3.32\!\times\!10^{-1}$  \\
0.75 & 0.03 & $6.98\!\times\!10^{-2}$ & $8.65\!\times\!10^{-3}$ & $2.91\!\times\!10^{-1}$ & $6.02\!\times\!10^{-2}$  \\
\hline\hline
\end{tabular}
\end{inlinetable}

Overall, we find that biasing the supernova detection probability according to 
explosion energy has a profound effect on the expected rate of observable 
black hole emergence. In general, we find that the event rate drops 
by a factor of 5-15 with respect to an unbiased detection probability, as 
depicted in Table~\ref{tab:navRATES}. Note that the effect is more pronounced 
for larger values of $f$, because the luminosity 
is driven by the energy that is not spent on unbinding the star, i.e., 
on $1-f$. The typical event rates with these revised assumption are mostly 
once per several years for $f_{\rm He}=0.01$ and once per tens of years for 
$f_{\rm He}=0.03$. Once again, a constant 
$M_{\rm ^{56}Co}=2 \times 10^{-3}\;M_\odot$ for all progenitors would be 
especially destructive in terms of decreasing the potential event rate. 

An additional source of potential bias that would reduce the 
observability of black hole emergence is luminosity due to interaction 
of the ejected envelope with circumstellar material (CSM). If the CSM is 
dense enough, a significant luminosity, $L_{CSM}$, 
will arise as the shock wave propagates through it. Interaction with the CSM 
is the preferred model for the subclass SNeIIn, which are characterized by 
a high, slowly decaying, luminosity after the recombination peak. 
In particular, \citet{ChevFrans94} demonstrated that a dense CSM could 
sustain a luminosity at a level of $L_{CSM}\approx 10^{40}\;$\ergss for 
several hundred years after the explosion. 
A luminosity at this level would black hole would obscure any potential 
accretion luminosity. Indeed, this was quite possibly the case for SN1994W 
\citep{Sollerman94W}, which otherwise could have been a promising candidate 
for observing black hole emergence.
\citet{Cappellaro97} estimated that Type IIn supernovae are relatively rare, 
and probably make up only $2-5\%$ of the total SNeII rate. However, they note 
that owing to the relatively high luminosities of these supernovae, in recent 
years their contribution to the {\it observed} rate of SNeII is $15-20\%$; 
their influence over the present study must therefore be examined. 
Moreover, since the CSM is essentially the residue of mass loss from the 
progenitor during the red giant phase, more massive progenitors are more 
likely to have produced denser CSM. 

While the rate of mass loss in the star's wind, $\dot{M}_{wind}$, is expected 
to increase with progenitor mass, many other factors will contribute to the 
overall properties of the wind and the resulting CSM. We note that 
there was no evidence of CSM interaction at any stage in the observation of 
SN1997D \citep{SN1997Dnew}. For the purpose of a quantitative estimate of the 
potential influence of CSM interaction on the prospects of 
observing black hole emergence, we assume that the probability of 
the accretion luminosity {\it not} being obscured by a high $L_{CSM}$ 
decreases for larger progenitors and scales as $M_*^{-\theta}$. We 
calibrate $\theta$ by imposing the following combination: assume that (a)  
significant $L_{CSM}$ occurs in $5\%$ of SNeII and that (b) if all 
supernovae with CSM interaction are observed, they make up $20\%$ of observed 
SNeII. Imposing these assumptions on the E-BIAS model described above,
we find that for $\gamma=2.7$ this combination is best reproduced with 
$\theta\approx 0.5$.
By including the effect of possible CSM interaction on the light curve 
according to this recipe and renormalizing 
the overall observed SNeII rate, we arrive at yet another set of event 
per observed SNeII fractions as a function of distance 
(hereafter the E+M-BIAS model).
These sets are presented in Figure~\ref{fig:DISBINBi}c. Note that these 
new fractions are naturally lower than the corresponding values in the E-BIAS 
model. The total predicted rates under these 
assumptions are given Table~\ref{tab:BiMERATES}. 

Quantitatively, the potential of CSM interactions contributes an additional 
effect of reducing the predicted event rates by $\sim 30\%$ with respect to 
Table~\ref{tab:BiERATES}. It also further reduces the event rate in the 
$\beta=8$ model with respect to the radioactive free one, emphasizing the 
significance of even a small abundance of \Coa in the lower mass progenitors. 
While the additional mass dependent bias is not as important as the explosion 
energy bias, owing to the different functional dependence on $M_*$, 
the effect of CSM interactions - as parameterized in our model - reduces the 
rate of observable black hole emergence to, at best, one per four years. 
For most of a parameter range, we conclude that with using HST capabilities 
and current limits of supernova searches this rate is probably not better 
than once per decade. 

\begin{inlinetable}
\def\arraystretch{0.8}%
\setlength{\tabcolsep}{5pt}
\caption{
Predicted rates (yr$^{-1}$) for observable black hole emergence 
in the model with an explosion energy and wind bias 
(see text - model E+M-BIAS). 
\label{tab:BiMERATES}}
\begin{tabular}{l l | c c c c}
\hline\hline
$f$ & $f_{\rm He}$  
& $M_{\rm ^{56}Co}=$ & $M_{\rm ^{56}Co}=$ & $\beta=2$ & $\beta=8$  
\\  &
&         0          & $2\!\times\! 10^{-3}\;M_\odot$
& (Eq.~[\ref{eq:M_Co}])
& (Eq.~[\ref{eq:M_Co}])\\
\hline
0.25 & 0.01 & $1.21\!\times\!10^{-1}$ & $7.49\!\times\!10^{-3}$ & $2.94\!\times\!10^{-2}$ & $9.73\!\times\!10^{-2}$  \\
0.25 & 0.03 & $3.95\!\times\!10^{-2}$ &     $\sim 0$        & $6.99\!\times\!10^{-3}$ & $2.85\!\times\!10^{-2}$  \\
0.50 & 0.01 & $2.16\!\times\!10^{-1}$ & $1.72\!\times\!10^{-2}$ & $6.86\!\times\!10^{-2}$ & $1.82\!\times\!10^{-1}$  \\
0.50 & 0.03 & $4.58\!\times\!10^{-2}$ & $3.72\!\times\!10^{-3}$ & $1.26\!\times\!10^{-2}$ & $3.55\!\times\!10^{-2}$  \\
0.75 & 0.01 & $2.84\!\times\!10^{-1}$ & $2.01\!\times\!10^{-2}$ & $1.22\!\times\!10^{-1}$ & $2.47\!\times\!10^{-1}$  \\
0.75 & 0.03 & $5.29\!\times\!10^{-2}$ & $6.52\!\times\!10^{-3}$ & $2.06\!\times\!10^{-2}$ & $4.48\!\times\!10^{-2}$  \\
\hline\hline
\end{tabular}
\end{inlinetable}

\section{Conclusions and Discussion}\label{Sect:End}

In this work we attempt a first estimate of the rate at which black 
hole emergence in supernovae may be observed. We used an analytic model and 
a parameter survey to examine the competition between the expected accretion 
luminosity and the other sources of power for the supernova light curve, most 
notably radioactive heating. In general, we show that the event rate is 
larger if the relative fraction of the total kinetic energy driving the 
expansion (not spent on unbinding the star) is smaller and if the fraction of 
this kinetic energy deposited in the helium layer (which is the source of 
late-time accretion) is also smaller. Furthermore, the 
event rate is larger for reduced amounts of radioactive heating in the envelope.  

With the efficiency of current supernova searches and the capabilities of 
the HST for detecting the accretion luminosity at emergence, we find that 
under optimistic assumptions, the rate of ``events'' -- successful 
observations of black hole emergence -- may be as large as a few events per 
year. If Type II supernovae in current searches are detected with no 
dependence on progenitor mass and explosion energy, then for most of our 
parameter range, the event rate should be better than once per two years. 
However, in the more likely event that detection of supernovae in current 
searches is weighted towards brighter supernovae, this rate is dramatically 
reduced. The more massive progenitors which lead to black hole formation are 
also likely to produce less luminous explosions, and would therefore be 
underrepresented in present searches. The best known potential candidate for 
allowing observable black hole emergence, SN1997D, was such a 
subluminous explosion. If a bias against discovering supernovae from 
explosions of more massive progenitors exists, it reduces the 
potential rate of finding black hole emergence in supernovae as 
well. We estimate that the event rate in the presence of such a bias is not 
larger than once per few years. We also find that some additional bias may 
arise due to obstruction of the accretion luminosity by interaction of the 
expanding envelope with circumstellar material. Such obstruction is more 
likely for the more massive progenitors which experience greater winds during 
their evolution; we view SN1994W as a likely candidate for potential 
emergence that was obscured by luminosity due to circumstellar 
interaction. We find that if probability of significant contribution to the 
light curve from interaction of the ejecta with circumstellar material 
is larger for more massive progenitors, the fraction of observable black hole 
emergence per observed Type II supernovae is reduced 
by an additional factor of $0.3-0.7$. In this later case the actual 
event rate for present detection capabilities and limitations of supernovae 
searches is unlikely to be greater than one per decade.

Our findings suggest that, at present, the prospects of observing black hole 
emergence in Type II supernovae are small, but not negligible. The immediate 
consequence is that if a low energy Type II supernova is discovered, high 
priority must be given to following it for several months until beyond 
recombination. If a small ($\sim 10^{-3}\;M_\odot$ or less) amount of \Coa is 
then identified in the tail of the light curve, it would be prudent to 
schedule HST observations of the declining light curve with hope of observing 
the emergence of the black hole. 

These rates might be somewhat increased by detailed 
spectral analyses, although a successful spectral analysis requires
a larger-than-threshold total luminosity. Since luminosity due to 
radioactive heating and CSM interaction is generated in the outer 
(and cooler) part of the expanding envelope, their spectra tend to be 
mostly in the R-band. Furthermore, in typical Type II supernovae the light 
curve is observed to become redder at later times, as the effective 
temperature of the emitting region decreases. By contrast, the accretion 
luminosity is generated deep in the ejecta, so its effective temperature 
should be that of partially ionized envelope material - 5000-10000 degrees. 
Frequency-dependent calculations of the accretion luminosity have not 
been carried out, but the intensity should peak in the visible or near IR.  
A light curve in which the R-band luminosity follows a decline rate expected 
from \Coa decay while the visible band luminosity appears to be decline more 
slowly could indicate the presence of the black hole even if accretion  
is still subdominant in the total bolometric luminosity.

Perhaps the most important conclusion of our analysis is that detectability 
of the supernovae themselves appears to be the dominant factor in the rate of 
observing black hole emergence, if indeed black hole producing supernovae are 
fainter than average. Clearly, a search for black hole emergence will benefit 
greatly from any dedicated faint supernova project, indirectly also from 
one aimed at high red shift explosions. If the bias we assumed against 
discovery of fainter supernovae (Table~\ref{tab:BiERATES}) can be at least 
partially lifted by such a search, we may be able to approach the higher 
rates found for an unbiased survey (Table~\ref{tab:navRATES}). An 
additional boost to the event rate would arise with a next generation 
instrument with improved capabilities. While \Cob and \Ti abundances may not 
allow detection of emergence luminosities lower than $\sim 10^{35}\;$\ergss, a 
more sensitive instrument will certainly increase the range over which higher 
emergence luminosities can be observed. In particular, we find that 
if the threshold apparent magnitude of the instrument can be improved to 
$m_v\sim 30$ (possible threshold for NGST), the expected event rate is 
increased by a factor of 6-8.  
  
In conclusion, we suggest that observing emergence of a black hole in Type II 
supernovae is potentially possible even with present capabilities. In view of 
the relatively low rate, high priority should be assigned if a likely 
candidate supernova (low energy, low \Coa) is discovered.  
The event rate would significantly enhanced with future improved 
instruments, and also through dedicated nearby faint supernova searches with 
present capabilities. Such a search based on ground instruments is currently 
being initiated (M.~Turrato 2000, private communication), and could also be 
included as part of other projects aimed at observing high redshift 
supernova. In view of the importance of achieving direct confirmation 
of the supernova -- black hole connection, we believe such efforts should 
be encouraged.

\acknowledgments

It is a pleasure to thank Luca Zampieri for many useful discussions and for 
important comments on the manuscript. We are also grateful to Chris Fryer 
for important discussions and for providing us with models of supernovae 
progenitors, and to Jesper Sollerman and Massimo Turatto for valuable 
communications. This paper was supported in part by 
NASA Grants NAG 5-7152 and NAG 5-8418 to the University of Illinois
at Urbana-Champaign.

\end{document}